AI-synthetic biology Convergence draft
Benjamin.d.trump@usace.army.mil

# The Convergence of AI and Synthetic Biology: The Looming Deluge

Cindy Vindman[1], Benjamin Trump[1], Christopher Cummings[1], Madison Smith[1], Alexander J. Titus[2], Ken Oye[3], Valentina Prado[4], Eyup Turmus[5], Igor Linkov[1]

[1]US Army Corps of Engineers; Benjamin.d.trump@usace.army.mil

[2]Information Sciences Institute, University of Southern California, Iovine and Young Academy, University of Southern California

[3]MIT

[4]Sphera Corp

[5]NATO Science Program

## Abstract

The convergence of artificial intelligence (AI) and synthetic biology is rapidly accelerating the pace of biological discovery and engineering. AI techniques, such as large language models and biological design tools, are enabling the automated design, build, test, and learning cycles for engineered biological systems. This convergence promises to democratize synthetic biology and unlock novel applications across domains from medicine to environmental sustainability. However, it also poses significant risks around reliability, dual use, and governance. The opacity of AI models, the deskilling of workforces, and the outdated nature of current regulatory frameworks present challenges in ensuring responsible development. Urgent attention is needed to update governance structures, integrate human oversight into increasingly automated workflows, and foster a culture of responsibility among the growing community of bioengineers. Only by proactively addressing these issues can we realize the transformative potential of AI-driven synthetic biology while mitigating its risks.

## 1. Introduction

Recent years have witnessed rapid progress in two transformative technological fields - artificial intelligence (AI) and synthetic biology. AI advances have been enabled by improvements in computational speed, data transfer, and data storage. Synthetic biology advances have been powered by improvements in reading, writing, and editing DNA. The use of AI in synthetic biology has evolved in two phased. Initially large language models (LLMs) and biodesign tools were used for biodesign. Now, machine learning is being employed to analyze integrated genomic and functional data sets. This convergence is yielding powerful discriminative assessments of biological information, system and structure which are accelerating, and democratizing bioengineering [1]. Current application of AI to biotechnical problem sets is delivering both rapid technological change and creating a deluge of new

AI-synthetic biology Convergence draft
Benjamin.d.trump@usace.army.mil

governance and oversight challenges. Future generative AI will likely deliver not just discriminative and predictive capability but perhaps an AI biological designer, cognizant and considerate of the contextual challenges presented by the biological domain. Responsible development of this AI-synthetic biology frontier necessitates proactive governance based on principles of knowledge cultivation, accountability, transparency, and ethics.

AI capabilities are facilitating a more complete understanding of biology, and this growing fluency will underpin AI-assisted biological engineering and eventually lead to a robust ability to imagine and validate a wide array of biological constructs. While we are still in the age of defining the tools and building materials that future synthetic biologists will employ, AI is now and will continue to hasten discovery and aggregation of biological information [2]. Efforts to deliver curated, intelligible biological information will eventually shift from discriminative to generative in nature, giving rise to automated bioengineering pipelines. Efforts like BioAutomata embody this vision, using AI to guide each step of a design-build-test cycle for engineering microbes - with limited human supervision [3]. The result could be dramatically accelerated and democratized synthetic biology [4].

However, this AI-synthetic biology convergence also poses risks if not developed thoughtfully [5]. Many of these risks are associated with a reduced knowledge threshold to carry out biological engineering tasks and the democratization of the tools and capabilities to engineer potentially harmful sequences or organisms of concern. Lack of oversight and access to emerging tools like desktop sequencers create potential scenarios where accidental or intentional de novo design of harmful biology is released and allowed to spread uncontrolled. The potential democratization of the design and testing of engineered biology could reduce our ability to anticipate the consequences of synthetic biological constructs. Further, the design and implementation of mitigation strategies for unforeseen consequences could move out of reach. There are also dual use issues if AI enables rapid production of harmful engineered organisms. More broadly, policy frameworks tend to lag cutting edge technologies, exacerbating the above risks within an environment of incomplete risk insight, and inconsistent policies across countries harden this challenge. While guidelines do exist for things like genome synthesis screening procedures, these are still merely recommendations and robust systems of oversight and transparency have yet to be mandated industry-wide.

Balancing these tensions inherent to AI-synthetic biology convergence requires multi-stakeholder collaboration and governance. Scientists, ethicists, policymakers, and other experts must work closely and transparently to ensure technologies advance responsibly. Biological construct design and deployment currently requires extensive regulatory oversight, that should continue to be the case, but a new question arises; If a design process, testing protocol, or deployment strategy happens in an increasingly distributed and automated manner what current governance instruments or regulatory protocols might be insufficient to gauge risk? Integrating oversight into highly automated pipelines could act as a safeguard to inform risk assessment, regulation, and policy, as could developing international soft laws and codes of conduct regarding safe use recommendations such as the screening and logging of synthesized DNA sequence [6]. If pursued judiciously, this nexus of breakthrough technologies could positively transform fields from human health to agriculture and environmental sustainability. To get there, we must thoughtfully weigh each step to understand points of sensitivity, intervention, and regulation in managing risks to balance safety with the economic first actor incentives for breakthrough innovation.

Ultimately, a balance must be struck between nurturing convergence capabilities to accelerate desirable breakthroughs across various product lines, and the imperative to diligently identify and appropriately govern novel risk on an international landscape. This article delineates the distinctive roles of AI in accelerating the design and experimental phases of synthetic biology. At the same time, we underscore the critical need to revisit regulatory requirements, instruments, and capabilities to ensure that risks stemming from technological convergence are adequately captured through pertinent hard or soft law for the coming decade. It is imperative to distinguish between the automation of routine tasks, which AI facilitates, and the decision-making processes that require human oversight and ethical consideration. The integration of AI into synthetic biology presents unparalleled opportunities for innovation yet necessitates a nuanced understanding of where automation serves to enhance efficiency and where human intervention is indispensable for ethical oversight and safety assurance. Specifically, this discussion examines the technical and ethical standards necessary for balancing automated systems with human-in-the-loop controls within AI-driven biotechnological pipelines. It also explores strategies for preserving critical human oversight in the design process, even as we advance towards more autonomous laboratory environments [7]. While AI significantly contributes to the field's advancement by optimizing design and experimentation, it does not obviate the need for rigorous regulatory frameworks or diminish the essential role of human oversight at various stages of the product lifecycle. As such, this article discusses some of the emerging opportunities and challenges stemming from the technological convergence of AI and biotechnologies like synthetic biology, while also suggesting key areas of attention and potential innovation to ensure effective but not excessively burdensome governance of technology risk.

## 2. Promises of AI-Synthetic Biology Convergence

The integration of artificial intelligence (AI) techniques into synthetic biology workflows is set to accelerate the design, testing, and optimization of engineered biological constructs across multiple domains [8]. From pharmaceutical production to environmental remediation, AI-enabled automation and in silico modeling can shorten development timelines and expand the complexity of achievable biosystems. Early efforts to incorporate advanced digital capability such as LLMs and BDTs foreshadow the near-term achievements of this convergence. Specifically, the non-trivial processing power of machine learning (ML), a data-driven subdiscipline of AI, will likely deliver rapid acquisition of complex high fidelity biological information, increasingly accurate sequence-to-structure prediction modeling and improved design-build-Test-Learn cycle efficiency. These advances will also be the foundation for future digital biodesign that will someday be capable of rapid automated design and synthesis of novel biological constructs ranging from macromolecules to entire metabolisms. Early examples of improvement of these aeras of bioengineering exist and we can use them as guideposts as we anticipate the future of AI empowered synthetic biology.

### Knowledge acquisition and refinement

The last 50 years have seen remarkable gains in the acquisition and interpretation of sequence information. Early sequencing technologies, based on chain termination, required hours of work by highly skilled hands to deliver short segments of nucleic acid sequence. These efforts eventually gave rise to the era of high-throughput sequencing marked by the introduction of automated sequencing platforms and the application of computer processing which delivered the assembly of the first long contiguous sequences of DNA [9]. Further ingenuity produced massively parallel sequencing techniques often referred to as "Next generation". DNA sequencing began to significantly outpace Moore's Law in 2008 This trend continues today [10].

Growing capability in sequencing empowered the human genome project as well as the delivery of multiple eukaryotic genomes in the early 2000s. The utilization of digital tools such as Basic Local Alignment Search Tool (BLAST) and microarray technologies encouraged the emerging fields of comparative -omics [11]. The success of the BLAST tools highlights how integral digital based processing is to modern biological investigation. This early convergence of a digital tool which allowed faster examination and delivered insights into the structure of both coding and noncoding sequence foreshadows the success of powerful AI/ML tools that will deliver the next generation of biological fluency.

Today, emerging single molecule sequencing (SMS) capabilities are delivering improved cost, speed, and platform portability. SMS is also opening the door to a more exquisite examination of sequence, the convergence of AI and SMS is producing more intricate sequence information including modified base calling, sequence variant calling, and chromosome phasing [12 - 13]. These advances have already delivered in application areas such as medical diagnostics, epigenomic analysis and the improvement of reference genomes [14 – 16]. In future it is likely that ML will allow raw data to be curated and interpreted to an even greater extent at the point of original discovery.

Molecular genetic studies have piece by piece revealed the intricate processes that control gene expression. The landmark identification and purification of the eukaryotic RNA polymerases in 1969 followed by decades of rigorous biochemical studies revealed a staggeringly complex interplay between DNA sequence, chromatin structure and the soluble factors that control the dynamic and responsive industry of eukaryotic gene expression. While impressive revelations regarding the paradigm of gene expression at large have been made, gaps in our ability to predict how both coding and non-coding genomic information deliver the dynamic living structures persist. As we continue to uncover the paradigms of genetic expression including nucleic acid sequence structure, epigenetic structure, and other contextual effectors; our understanding of how biological function is recorded, stored, and altered will grow more sophisticated. Fulsome cognizance of how biological information is transformed into functionality is almost certainly unobtainable without the aid of the analytical power of AI. AI empowered analyses of the biological systems may themselves fall short of this immense task, but they will move us closer to mastery.

Tools like LLMs and BDTs, and other technologies that broadly fit under the umbrella of AI, are beginning to help shape our understanding of genomic information. AI is being employed to progress our understanding of how genetic sequence becomes physical structure. Significant capability is emerging in DNA-based LLMs and BDTs that are capable of tasks such as gene finding, enhancer annotation and chromatin accessibility prediction. Ultimately, this convergence will enhance human understanding of how biological structures are produced in a temporally and spatially coordinated manner to produce functional metabolisms.

## Predicting Functionality

Beginning with the publication of the central dogma, perhaps the birth of modern molecular genetics, molecular biology has pushed us towards a better understanding of how stored biological information is transformed into structural capability. With every gain we seem to uncover a better but more daunting view of the intricate and sophisticated biochemistry which delivers the diversity of life on earth. In the section above we discussed the emerging convergent technologies that will allow for extensive and precise readings of sequence, and it is important to note this activity as a foundation for what we will discuss in this section. The integration of huge volumes of genomic data into more than nonsensical letters has, for decades, been a burdensome task. Modern molecular biology has partially revealed the significance of non-coding sequence, epigenetics, and other contextual effectors of biological manifestation. Adept synthetic biological designers will require mastery over a polyfactorial system, which is not fully understood. AI/ML supported knowledge acquisition will progress human understanding of the relationship between sequence, context, and structure.

Modern AI, including LLMs and BDTS, will be powerful tools in the deciphering of DNA data that will unquestionably improve our understanding of genomes and their design paradigms. AI is fueling advances across the biological sciences, from deciphering the rules of protein folding to optimizing chemical synthesis pathways [17]. AI, driven by state-of-the-art architectures like Transformers and Hyena models[4], are emerging as increasingly reliable tools for uncovering subtle, distant, and non-obvious implications of coding and non-coding sequence. The work to decipher the meaning of genomic data is more challenging than similar work on protein sequence, hindered largely by a lack of well-curated and publicly available experimental data. This discrepancy is caused by the fact that protein sequence has already been extensively experimentally decoded, removing the myriad intricacies of expression altering non-coding detail that is abundant in genomic data. Put another way, the derivation of phenotype from DNA sequence will require deeper understanding of a language that has been developed and refined via 4 billion years of evolutionary process.

Proteins are the molecular workhorses of life, and the physical result of the central dogma. Protein function is derived from its exact physical embodiment, understanding how a protein will play its metabolic role requires intricate awareness of its shape and charge to the atomic level. Empirically derived 3D protein structure has historically required laborious techniques such as X-ray crystallography that placed some proteins, including many membrane-bound proteins, out of reach for structural biologists. In the first decade of the 21$^{st}$ Century cryo-electron microscopy improved the plight of structural biologists by removing the need for crystallization prior to molecular interrogation [18]. It is still however, a non-trivial task to identify precise physical 3D structure of proteins.

Protein engineering stands to benefit from AI [19]. One key area of focus has been the de novo design of proteins - creating novel protein sequences predicted to fold into desired shapes and functions. Recently, DeepMind's AlphaFold has solved a 50-year-old challenge that has stumped the field by achieving improved accuracies at modeling protein tertiary structure for all known proteins simultaneously [20]. It is not surprising that in 2022 Nature Methods identified the AlphaFold2 protein structure prediction as the Method of the Year [21]. This computational leap forward approaches the level of accuracy of traditional empirical methods but does so for all known proteins simultaneously and delivers results with a significantly reduced time, cost, and labor burden [22]. AlphaFold2 has flexed its capability, predicting protein structure for all of the known human proteome [23]. Generative models such as Hyena [24], or from companies like Absci and Orbit Discovery use their in-house AI to propose novel protein sequences tailored for binding affinity, catalysis, signaling functions, and others. These AI techniques enhance rational protein engineering efforts and put in reach combinatorial spaces too vast for high-throughput screening. By exponentially accelerating the design proposal and selection, they stand to unlock novel biomolecules for applications from industrially useful enzymes to living therapeutics.

## Accelerating Design Cycles and Automating DBTL

Grueling, painstaking work gave rise to our first understanding of gene expression. The lac operon was explained by Nobel laureates: In 1969, Jacob and Monod [25] began to uncover the machines responsible for eukaryotic gene expression. The first wave of foundational discoveries regarding natures control over transformation of information into physical structure have since been joined by a myriad of molecular mechanisms such as small interfering RNA (siRNA) and clustered regularly interspaced short palindromic repeats (CRISPR) [26], the latter of which earned the authors the Nobel Prize in 2020. The foundational work in molecular genetics, while exciting, has also revealed the incomplete status of our understanding. That same insight revealed the daunting and complex challenge for the field of molecular biology. AI techniques, such as traditional machine learning algorithms and the more recently developed language models (LMs) and biological design tools (BDTs), are beginning to assist us in answering that challenge. As Science delivers a deeper understanding of how biological language is translated into physical structure, the toolkits of future synthetic biologists are being built.

The field of synthetic biology is approaching a tipping point driven by the application of ML [27]. Revolutionary ability to augment and automate computational steps in the design-build-test-learn pipeline will be delivered by AI [28]. For DNA design, neural network models may learn to optimize regulatory sequence and expression regimes for a desired biological context [29]. Sophisticated models can even propose entire genetic circuits for a specified outcome. Companies like Ansa Biotechnologies, TeselaGen, and Synthace offer such AI-guided DNA design and optimization services to clients engineering microbial strains or developing gene therapies. An entire industry of design optimization for the user of synthetic biological structure is emerging.

Engineered CAR T-cells have shown effectiveness against some lymphomas. These treatments are expensive, costing several hundred thousand dollars. This price point is a function of the effort required to design and implement the production of CAR proteins in the patient's T-cells. Further, while these can be effective customized therapies, they continue to have major limitations such as off-target toxicity. As databases of CAR-T designs are built researchers will begin to piece together a wider understanding of why certain constructs are effective. The application of ML to predict the quality of the complex interactions between CAR-T cells and their cancerous targets has been shown to track with clinical outcomes for an existing CAR-T cell treatment. The continued application of ML and future AI systems with access to growing databases will feed the ability of AI to predict functionality, ultimately lowering the bar for delivering efficacious, financially obtainable, individualized therapies.

Beyond construct design, AI can also automate and enhance downstream steps like molecular cloning, strain engineering, phenotypic assays, and data analytics. Robotics controlled by algorithms handle material transport, instrumentation control, colony picking, liquid handling, incubation, and chromatography. They can systematically build genetic variant libraries, perform multiplexed experiments, and characterize engineered cells with minimal human intervention. Startups like Biotium, Strateos, and Emerald Cloud Lab already leverage such capabilities, offering services like rapid microbial strain and enzyme optimization to clients. The automated build-test loops they orchestrate help engineer organisms for goals from biosensing to biomanufacturing.

Closing the build-test loop, AImay be useful in digesting and learning from the resultant data. Beyond accelerating each individual step, algorithmic coordination also continually tunes the end-to-end pipeline. Performance metrics from assays and analytics further refine design parameters, DNA synthesis constraints, robotic workflows, and models themselves. BioAutomata, an automated robotic platform coupled with predictive ML was able to demonstrate optimization of lycopene production pathway. This example removes the human in the loop after initial query, returned a completed DBTL cycle and delivered impressive optimization while testing less than 1% of variants (HamediRad et al., 2019).

Connecting AI analysis to automated empirical learning will allow rapid interrogation of synthetic design across a spectrum of cellular and multi-cellular contexts. While this will clearly reduce costs and labor input required to identify functional synthetic biological constructs it's also worth noting that it reduces human access to empirical knowledge acquisition.

## Enabling Novel Biosystems

Beyond sheer acceleration, AI integration can also expand the complexity frontiers of achievable biological systems. Tasks like controlling and interpreting multiplexed sensors, tuning multidimensional gene expression, or optimizing intricate metabolic pathways require assessing vast design spaces. Computational exploration of combinatorial and sequence spaces facilitates the rational design of multifaceted systems previously out of reach. For example, companies like Lycia Therapeutics and Nuvai leverage generative neural networks to engineer novel protein machines, signaling modulators, and smart enzyme cascades.

Synthesizing such elaborate blueprints demands a fluency in biology's design grammar - understanding how low-level DNA syntax translates to high level systemic functions. Here too AI is proving adept at deducing underlying design rules. Whether by mining patterns in databases or learning sequence-structure-function mappings from laboratory data, algorithms uncover predictive models relating genotypes to phenotypes. In a feedback loop, experimentally validating model outputs also continually refines understanding of this grammar. The design of an optimized whole gene regulatory structure using a deep generative adversarial network can be used to drive regulatory control above traditional mutagenesis methods. Startups like Design-by-Data and Flatcarbon leverage such learned design principles for forward engineering of microbes, yeast, or cell lines to specification.

As algorithms are engineered for improved interpretability of genetic information at a biological system level, they can assist bioengineers in consciously composing increasingly sophisticated systems for sensing, manufacturing, remediation, and medical needs. Rather than just troubleshooting known designs via discriminative models, these AI systems will become generative partners enabling more expansive and reliable creation. Ultimately AI will deliver a next generation artificial bio designer. An AI biodesigner will require a more sophisticated ability to apply the polyfactorial contextual effectors that lie between nucleotide structure and biological function to the task of bioengineering. The advent of this AI biodesigner will be a leap forward from the current discriminative assistance that is currently in use. Progress towards a capable AI biodesigner must be accompanied by human knowledge capture, critical for both installing appropriate interrogation sites and controls on next generation biotechnical AI models.

## Increased Access and Reducing Skill Threshold

The convergence of AI and synthetic biology is poised to dramatically lower the skill threshold allowing access to and participation in the bioengineering landscape (O'Brien & Nelson 2020). By automating routine molecular biology tasks and providing intuitive design tools, AI lowers the barriers to entry and de-skills many routine technical tasks for a wider range of interested actors. Traditionally, the field has been restricted to highly skilled experts with extensive hands-on experience in molecular biology techniques. However, the integration of AI is now enabling computer scientists, entrepreneurs, and even biohackers to engage in bioengineering projects with minimal wet lab backgrounds.

One key way AI facilitates this democratization is by handling repetitive workflows and providing user-friendly interfaces. Graphical user interfaces (GUIs) abstract away the complexities of command-line programming, allowing those without coding expertise to still leverage advanced models. Startups like Strateos and Emerald Cloud Lab take this a step further, offering remote access to robotic instrumentation for automated experimentation. This means even freelance bioentrepreneurs can prototype ideas without the need for costly in-house lab infrastructure.

Moreover, as AI models grow increasingly sophisticated, they are beginning to encapsulate the domain knowledge and decision-making capabilities that were once the exclusive purview of seasoned researchers. By codifying the heuristics and intuition of human experts into algorithmic routines, AI is progressively deskilling certain aspects of the bioengineering process. In the near future, AI assistants may provide personalized guidance and support, enabling students, DIY scientists, and citizen synthetic biologists to safely explore ideas without direct supervision from established practitioners.

However, it is crucial to recognize that this democratization also comes with inherent risks. As the tools and knowledge required to engineer living systems become more widely accessible, so too does the potential for accidental or deliberate misuse. While AI can streamline technical workflows, it cannot replace the ethical judgment and social responsibility of human actors. Therefore, appropriate safeguards, oversight mechanisms, and educational initiatives must be put in place to ensure that biosafety and biosecurity standards are upheld even as the field expands to welcome new participants.

## 3. Risk and Governance Challenges

Along with the benefits discussed above, the integration of AI into synthetic biology also creates risks related to reliability, dual use, and outmatched governance systems. Unlike more bounded applications, programmed cells can self-replicate, evolve, and disperse with ecological consequences at stake. Employing emerging AI capabilities to engineer organisms demands heightened safeguards and oversight. Core areas needing scrutiny include opaque AI models, automation reducing human diligence, potential for weaponization, and outdated regulations [30].

### Interpretability of AI Models
Many AI models for biodesign like generative neural networks or gradient boosting models operate as "black boxes" - delivering predictions without explanations for internal reasoning [31 - 32]. While this opaqueness does not hinder their technological utility, it does limit evaluability regarding reliability or safety and may also retard acceptance and legitimization of AI models for biodesign. For instance, a protein design large language algorithm may hallucinate flawed sequence suggestions that nevertheless receive high performance scores. Benchmarking new AI techniques against traditional expert methods can give insight into the relative limitations of these techniques regarding certain tasks, but still cannot capture the detailed reasons why a model reaches a specific conclusion[33 – 34]. Without transparency into failure modes, researchers cannot fully trace or troubleshoot limitations, this lack of insight may give potential adopters pause and erode confidence even for those tools that demonstrate functionality.

The difficulty in understanding the intent behind AI-generated outputs poses policy challenges. While AI models may be highly accurate in their predictions, they are ultimately a reflection of the data they are trained on. If the training data incorporates biases, either from the underlying biological systems or from the human curators, these biases can propagate through to the model's outputs, which then go on to influence policy interpretations as well as our understanding of human and environmental safety [35]. Moreover, even if a model correctly identifies patterns or relationships in the data, it may not capture the proximate causes or mechanistic explanations for these associations. This lack of causal understanding limits the ability to anticipate potential side effects or failure modes when translating AI predictions into real-world applications.

Interpretability goes beyond mere prediction - it involves understanding the meaning, value, and justification behind a model's outputs. Factor analysis techniques can help uncover the latent variables driving a model's decision-making process, providing a reduced functional form that is more amenable to human comprehension [36]. By examining the differences between structure and function learned by the model, researchers can gain insights into the biological mechanisms underpinning its predictions.

This interpretability is crucial for validating the claimed benefits of AI-assisted biodesign while also identifying potential risks. Understanding how a model maps from training data to findings to ultimate justifications allows for more rigorous evaluation of its real-world applicability. Techniques like saliency maps, counterfactual explanations, and feature importance rankings can help illuminate the key factors influencing a model's outputs [37]. Armed with this knowledge, domain experts can better assess whether a model's reasoning aligns with established biological principles and experimental evidence.

Achieving interpretability remains a challenge, particularly for complex models operating on high-dimensional data. The sheer number of parameters and non-linear interactions can make it difficult to distill a model's decision-making process into a form that is easily digestible by humans. Moreover, there may be inherent trade-offs between model performance and interpretability, as the most accurate models often rely on intricate architectures that resist simple explanations (Murdoch et al., 2019).

Moreover, biosecurity risks grow if algorithms have undetected flaws or training biases hackers can exploit to deliberately output hazardous designs. DARPA's recent malicious AI report war gamed scenarios around poisoning data or models for biomanufacturing, highlighting vulnerabilities of opaque systems. For any AI-bio convergence, standards requiring explainability, auditability, and transparency into variables influencing output would bolster accountability and trust. Alongside monitoring for signs of data or model tampering.

The ability of AI to autonomously interpret and respond to observed threats is limited but developing. AI's ability to self-teach and solve problems in synthetic biology extends beyond human capabilities, largely due to its proficiency in handling and analyzing vast datasets. AI can identify patterns and relationships in genetic data that are too subtle or complex for human researchers to discern. This leads to the identification of problems that humans might not have recognized or understood how to address. For example, AI might discover non-obvious genetic interactions that influence drug resistance in pathogens, a problem that human scientists might not have identified due to the complexity of genomic interactions. In industrial enzyme development, AI could reveal new enzymatic pathways that optimize production processes, pathways that human researchers might have overlooked due to the sheer volume of potential enzymatic combinations and reactions.

The implications of this capability are profound. AI-driven discoveries can leapfrog current scientific understanding, but they also present challenges in terms of verification, safety, interpretation, and ethical considerations. The advanced solutions proposed by AI might be effective, yet their underlying mechanisms could be opaque, making it difficult to predict long-term effects or unintended consequences. This opacity necessitates new frameworks for risk assessment and management in AI-assisted synthetic biology innovations, balancing the potential for groundbreaking advances with the need for safety and ethical responsibility [38].

Further, the interpretation of AI models is not purely a technical matter, but also involves subjective values and cultural contexts. What is considered a desirable or acceptable outcome may vary widely across different countries and communities. For example, the use of AI to optimize gene drives for environmental conservation might be viewed favorably in some regions, while others may prioritize preserving natural ecosystems without human intervention [39]. Balancing these competing values and priorities requires inclusive deliberation and participatory governance that goes beyond the capabilities of AI alone. As such, policymakers must grapple with the challenges of regulating a technology that is rapidly evolving, difficult to interpret, and entangled with broader societal concerns that include competing incentives, perspectives, and interpretations of risk across national borders.

### Ensuring Human Oversight in AI-Automated Workflows

A critical concern as AI assumes greater responsibility for biological design, building, and testing is maintaining adequate human oversight to identify and mitigate potential risks [40]. The increasing automation of workflows may lead to the deskilling of workforces, as personnel become overly reliant on algorithms without critically evaluating their suggestions or outcomes. This lack of human vigilance could allow unsafe engineered organisms to slip through automated build pipelines. The COVID-19 pandemic, for instance, exposed gaps in screening protocols for emerging viral sequences from genomic databases [41]. Such incidents highlight the ongoing need for human diligence and oversight, even when working closely with advanced AI tools.

To address this challenge, bioengineers must establish ethical standards and protocols that keep humans in the loop at critical assessment points as research pipelines progress. The specific functions and decision gates requiring human evaluation will vary depending on the context and safety considerations. However, it is crucial to develop clear guidelines that delineate the roles and responsibilities of human experts in validating AI-generated designs, monitoring experimental outcomes, and making hard calls when faced with limited transparency or intelligibility of models.

Moreover, as AI capabilities advance, we may soon witness the emergence of fully autonomous biodesigner that can handle the entire process from initial query to final construct delivery. While such AI-driven platforms could revolutionize the field, their development must be accompanied by thoughtful construction of human-in-the-loop regimes. These oversight mechanisms should be enforced through professional norms, funding requirements, and regulatory frameworks to ensure that AI-assisted bioengineering remains accountable to societal values and priorities.

Importantly, the integration of AI into bioengineering workflows should not be viewed as a replacement for human expertise, but rather as a tool to augment and enhance human capabilities. Ongoing workforce training around the responsible development and ethics of converging technologies will be essential to counteract deskilling risks and ensure that researchers can effectively leverage AI while maintaining critical thinking skills [42]. By fostering a culture of continuous learning and ethical reflection, the field can harness the power of AI-automated workflows while safeguarding against unintended consequences.

## Dual Use Potential

Even with responsible practices in place, the possibility of misuse by state or non-state actors cannot be entirely eliminated [43]. For example, automated DNA synthesis platforms controlled by algorithms could be covertly manipulated to generate pathogenic sequences or optimize the virulence of existing pathogens. While such biosecurity risks predate the emergence of AI-synbio convergence, the accelerated pace and expanded scope of bioengineering enabled by these technologies can strain existing governance and security frameworks [44]. Ultimately, there are emerging concerns that AI-biotechnology convergence may inspire or amplify dual use research of concern (DURC) ("research that can be reasonably anticipated to provide knowledge, information, products, or technologies that could be directly misapplied to pose a threat with broad potential consequences to public health and safety, agricultural crops and other plants, animals, the environment, materiel, or national security)" [45].

Moreover, the dual use potential extends beyond the direct synthesis of pathogenic agents. AI-assisted bioengineering could also be used to enhance the transmissibility, stability, or target specificity of existing pathogens, leading to the creation of novel threats. Techniques such as directed evolution and gain-of-function (GOF) research, which can involve modifying pathogens to increase their virulence or host range, are particularly concerning in this regard [46 – 47]. While such research can provide insights into pathogen biology and inform the development of countermeasures, it also carries inherent risks of accidental release or deliberate misuse.

The digitized and distributed nature of AI models and tools further complicates efforts to prevent misuse. Unlike physical materials, digital files containing AI algorithms or DNA sequences can be easily shared and replicated across borders, making it difficult to track and control their dissemination [48 – 49]. Moreover, the increasing accessibility of DNA synthesis technologies means that even non-experts can potentially create novel biological threats using AI-generated designs.

To mitigate these risks, a multi-pronged approach is needed that encompasses both technical solutions and policy interventions. From a technical perspective, enhanced screening methods are required to detect and filter out potentially dangerous sequences, including those generated by AI algorithms. However, existing sequence-based controls may struggle to identify novel or artificially designed sequences with unpredictable functions [50]. Developing more advanced screening technologies that can assess the functional characteristics of DNA sequences, rather than relying solely on homology to known pathogens, will be critical.

On the policy front, international coordination and harmonization of governance frameworks areneeded. Currently, norms and regulations around DNA synthesis and dual use research vary widely across countries, creating gaps that can be exploited by bad actors [51]. Establishing global standards for transparency, supply chain tracking, and information sharing can help create a more robust and responsive biosecurity ecosystem. There is growing recognition of the need to address dual use and biosecurity challenges in the context of AI-synbio convergence. International organizations such as the World Health Organization (WHO), INTERPOL, and the National Institute of Standards and Technology (NIST) have recently highlighted these issues and called for proactive policy measures [6, 52]. However, translating these high-level recommendations into concrete and enforceable policies remains a work in progress.

Preventing the misuse of AI-assisted bioengineering will require a sustained and collaborative effort from researchers, policymakers, security experts, and civil society. By proactively addressing dual use risks and investing in responsible innovation frameworks, we can work to ensure that the potential of these technologies is realized in service of the greater good, rather than being subverted for harmful purposes.

## Regulatory Shortcomings – the Pacing Problem Anew

Finally, the cross-cutting risks from AI-synthetic biology integration highlighted above also expose governance gaps - as regulations struggle catching up to fast changing technological capabilities. Few existing policy frameworks contemplated risks around autonomous generation of digital genomic blueprints or sequence-based controls for now widespread custom DNA synthesis abilities. And oversight bodies like the Recombinant DNA Advisory Committee in the US face criticism for lacking binding rule making authority, transparency, and cultural competency surrounding new sciences.

Managing the rapid pace of technological change, termed the "pacing problem", poses an endemic challenge for governance systems. Policymaking inherently moves slower than exponential tech advancement – even in instances where policy priorities desire rapid modernization of technology capabilities [53]. This is particularly evident in the rivalry between the United States and China, which are both heavily investing in AI and biotech research as part of their broader geopolitical strategies [54]. This competition has spurred investments in research and development, as well as efforts to attract top talent and gain access to sensitive technologies. China has made no secret of its ambitions to become a global leader in AI and biotech, with the government launching a series of initiatives and funding programs to support these goals. For example, the "Made in China 2025" plan identified biotechnology as a key strategic industry, while the "New Generation Artificial Intelligence Development Plan" outlined a roadmap for China to achieve dominance in AI by 2030 (Kania, 2020). The United States, for its part, has responded to China's challenge by ramping up its own investments in AI and biotech research. The National Defense Authorization Act for Fiscal Year 2021 included provisions for a new National Artificial Intelligence Initiative [55]. Competition between these and other partners carries profound implications for economic growth, defense, and health, with rewards incentivized towards as 'first actor privilege' [56].

With accelerating innovation comes the struggle for risk informatics and risk governance to 'keep up'. This challenge is not restricted to AI-biotechnology convergence, although the pace of such convergence raises an array of near-term and long-term technology governance questions. This lag leaves gaps where innovations progress absent oversight, sometimes enabling unanticipated harm or coercion before safeguards are activated. The dilemma grows as technologies like AI and synthetic biology converge, integrating powerful capabilities faster than risks are characterized or governed. Unfortunately, the dilemma will be dwarfed with the advent of an artificial biodesigner.

For example, DNA synthesis and gene editing techniques are now rapid, inexpensive, and accessible thanks to technological advancement [57]. Yet screening policies fail to match the volume of users. This means oversight depends largely on voluntary self-governance - hoping actors internally weigh benefits and risks. But such self-policing falters securing collective interests against errors, externalities, or malicious non-compliance. Accordingly, the lag between technological possibility and prudential control widens, allowing potential slippage. The lag further widens with integrated AI-synthetic biology capabilities automating design and build cycle but consider the regulatory implications of an artificial biodesigner that only requires a query and access to automated wet lab capability to deliver an optimized structure. The advent of an artificial biodesigner will require a regulatory regime that considers data access, forced human in the loop safety and functionality reporting and restrictions on access to automated wet labs. These necessary controls will require interfacing across multiple industries and agencies to ensure chain of custody like consideration of the artificial design process.

All dual use technologies wrestle with this pacing challenge, but AI and synthetic biology feature acute attributes rendering governance uniquely difficult. These include hyper-scalability enabling systems to quickly disseminate globally once built, uncertainty given synthetic biology's complexity and AI's 'black box', and dual use traits innately embedded directly into underlying knowledge itself rather than just artifacts. Once digitized, information spreads rapidly and indefinitely. These facets distinguish bio or cyber risks from nuclear, accentuating policy lags.

Creative solutions are needed to address the pacing problem for AI-synthetic biology integration, lest unmanaged divergence erodes safety. Options range from anticipatory governance models that forecast and pilot policy ahead of full deployment to responsive capacities via global monitoring for risky convergence signals. But at core, rapidly modernizing legal frameworks via international technical resources, participatory review boards to calibrate oversight, and adaptive policymaking tools provide foundations [58]. With vigilance and collective responsibility, the pacing problem, while impractical to fully solve, can at least be mitigated [59].

Globally, provisions around transparency, licensing for restricted techniques like gene drives or live research demonstrations vary widely between countries and institutions [60]. Voluntary codes of conduct similarly exhibit little uniformity, compliance verification, or enforcement teeth industry-wide. The deficiencies of self-governance models grow starker amidst military investments, commercial secrecy imperatives, and global tech rivalries around domains like AI, synthetic biology, quantum, and robotics. Thus, improved governance guardrails and international partnerships appear essential to help sustain tech innovation that enhances collective well-being rather than eroding it.

## 4. First Steps to Demystify the Black Box

Realizing the transformative potential of AI- synthetic biology convergence is contingent upon navigating the complex landscape of benefits, convergent risks, and governance. Proactive governance and cooperative efforts among key stakeholders become the linchpin in harnessing the positives while responsibly mitigating the associated risks. While there are many steps required to improve the collective technologies' governance, a critical first step for many nations is the urgent need to address

AI-synthetic biology Convergence draft
Benjamin.d.trump@usace.army.mil

the AI-synthetic biology black box and identify opportunities to make the convergence learning and improvement process one that is traceable, defensible, and informed by institutional norms and values. These concerns coalesce into a central policy challenge shared by nations worldwide: existing governance tools are currently grappling with the present forms of AI, primarily centered on enhancing rather than fundamentally altering synthetic biology research and operations. However, as we look ahead to the next ten to twenty years, these instruments may prove insufficient in overseeing AI-driven synthetic biology. The imminent advancements in technology have the potential to transform or even supplant human involvement in learning, intuition, and the steering of synthetic biology research through automation and self-learning. While it is difficult to discern what specific governance strategies or instruments are needed to address near future AI convergence capabilities, a few hints have already emerged at some of the directions that policymakers and synthetic biology stakeholders might consider.

At present, the potential lack of explainability of AI-synthetic biology outputs, as well as the but uncertain implications to human health and biodiversity, are critical shortcomings that may stymie future development in key development areas such as medicine or environmental remediation. Demystifying the AI-Synthetic biology black box necessitates involving a deep integration of human intelligence at critical junctures of AI-driven processes, bolstering safety and security frameworks, and laying down transparent, actionable pathways for regulatory bodies and stakeholders to scrutinize AI's methodologies in interpreting genetic data and forging new biological innovations. Absent human guidance, the recursive AI learning process can generate potential opportunities for biological and genetic breakthroughs, yet equally could struggle with unforeseen errors in training data, or even learn from hallucinated interpretations of training data. This concern is not unique to AI's convergence with synthetic biology, though the implications for error are potentially more concerning, and can yield irreversible, sweeping harm to those exposed to AI-generated synthetic biology innovation.

Many of the potential benefits and risks discussed in this article relate to the challenge of demystifying the AI-synthetic biology black box. On the positive side, AI tools like large language models and biological design tools can help uncover subtle patterns in vast genomic and biological datasets, accelerating scientific understanding and the development of beneficial applications in medicine, agriculture, and environmental protection. Putting AI's predictive power in the hands of a wider range of users through automated labs and intuitive interfaces could democratize problem-solving and unlock innovative solutions from diverse contributors.

However, the opacity of many AI models and their use of underlying training sets makes it difficult to interpret how they are arriving at design recommendations or intervention strategies. Coupled with the deskilling of workforces and the abstraction of laboratory work into black-box machines, this opacity risks scenarios where unsafe or improperly vetted biological entities are created without adequate oversight. The dual-use potential of engineered organisms developed through such opaque pipelines further compounds the risks. Shortcomings in our ability to screen for concerning genomic sequences or assess emergent functions means questionable research could proceed unchecked.

The integration of 'human-in-the-loop' systems serves as a foundational pillar in iteratively interpreting AI-synthetic biology output, and improving transparency and explainability in how the AI biodesigner makes sense of opportunity, risk, and the most efficient paths forward to drive further innovation from basic science to commodifiable product (Figure 1). By strategically positioning domain experts within the AI decision-making workflow, regulators and policymakers can ensure a continual oversight mechanism that leverages human intuition and ethical judgment to guide AI's exploration of genetic codes and biological systems. This approach not only anchors AI's computational capabilities within a framework of human values and ethical considerations but also enhances the reliability of outcomes by incorporating expert feedback to refine algorithms and correct course as necessary. The dynamic interplay between human oversight and AI's processing power is critical in identifying and addressing biases, improving training data, evaluating hallucinations, ensuring ethical compliance, and validating the scientific integrity of AI-generated hypotheses and designs.

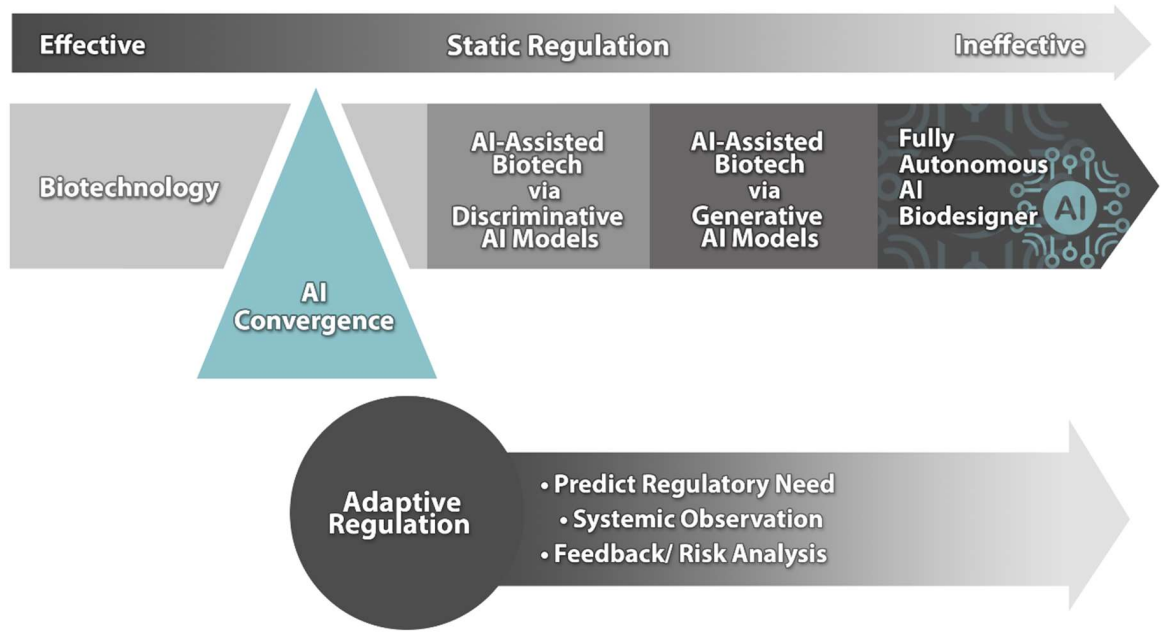

Figure: Effective Regulation for Convergence of Artificial Intelligence and Synthetic Biology. Over time, gaps between existing relatively static regulatory mechanisms and challenges associated with the convergence of AI and genomics are likely to increase. Effective oversight over emerging biotechnology processes and products will become increasingly problematic. Adaptive regulatory systems, featuring systematic observation and feedback, will be better able to respond to challenges posed by the convergence of AI and genomics than either static or anticipatory regulatory systems. Uncertainty over evolving technologies, applications and implications will undercut the viability of forecasting.

Balancing the benefits and risks of the AI-biotech convergence ultimately comes down to implementing the appropriate governance frameworks and oversight mechanisms. Central to this is integrating human judgment and accountability at key chokepoints in increasingly automated discovery and development workflows. While it may slow the pace of innovation, this is a necessary brake to avoid unintended and potentially catastrophic consequences. Domain experts must be in-the-loop to contextualize AI outputs, watch for failure modes, and make hard judgment calls - even if the underlying models are not fully transparent. Simulation sandboxes can further aid in pressure-testing AI-generated hypotheses before real-world deployment.

AI-synthetic biology Convergence draft
Benjamin.d.trump@usace.army.mil

Clear and enforceable guidelines are needed around acceptable use cases, containment protocols, monitoring requirements, and other safeguards. These should be developed through multi-stakeholder dialogues to navigate complex ethical quandaries and ensure alignment with societal values. Consistent standards for data access, model documentation, and impact assessment can improve auditability. And robust horizon scanning for emerging risks and ongoing public communication are critical for staying ahead of the pacing problem.

A likely first step towards identifying insertion points for human-in-the-loop involves the development of advanced analytical frameworks and simulation environments. These environments decision support algorithms and neural networks to simulate the intricate dynamics of biological processes, such as enzyme-substrate interactions, gene expression patterns, and cellular metabolism. The specificity and accuracy of these simulations are enhanced through the incorporation of vast biological databases and machine learning models that have been trained on genomic, transcriptomic, proteomic, and metabolomic data. This allows for a granular level of simulation fidelity, where even minor perturbations in genetic sequences or environmental conditions can be analyzed for their downstream effects on biological systems. Human experts, by interacting with these simulations, can apply their domain-specific knowledge to evaluate the feasibility of AI-generated predictions, scrutinize the models for potential biases, and ensure the simulations adhere to established biological principles. By simulating complex biological systems, these environments allow for the testing of AI-generated hypotheses and interventions in a controlled, limited, virtual space before any real-world application. This setup enables human experts to iteratively evaluate and refine AI's predictions, ensuring that the outputs are not only scientifically plausible but also ethically and socially acceptable. In turn, such inquiry can help identify areas where human intervention is necessary or desirable to address various limitations or concerns of a larger biodesigner.

The establishment of clear, transparent pathways for regulators and key stakeholders to evaluate the inputs, processes, and outputs of AI systems in synthetic biology is another cornerstone in addressing the black box challenge. This entails the development of standardized metrics and benchmarks for assessing AI's performance and reliability in biological applications, coupled with the creation of open-access repositories for AI-generated data, models, and findings. Such measures not only facilitate rigorous, independent verification of AI-driven innovations but also promote an ecosystem of accountability and trust among researchers, practitioners, and the public. Engaging regulatory bodies early in the development cycle and ensuring their active involvement in shaping the ethical and governance frameworks around AI in synthetic biology are essential steps in aligning technological advancements with societal norms and regulatory standards.

No amount of governance will be able to completely eliminate all risks. But by proactively grappling with these challenges, we can strive for a net positive impact - where transformative breakthroughs that improve the human condition outweigh the unavoidable missteps and growing pains. Getting this balance right is daunting but existentially important for future technology innovation.

AI-synthetic biology Convergence draft
Benjamin.d.trump@usace.army.mil